# Direct Reuse of Aluminium and Copper Current Collectors from Spent Lithium-ion Batteries

Pengcheng Zhu [a, c], Elizabeth H. Driscoll [b, c], Bo Dong [a, c], Roberto Sommerville [b, c], Anton Zorin [b, c], Peter R. Slater [a, c], Emma Kendrick [b, c] *

[a] School of Chemistry, The University of Birmingham, Birmingham, B15 2TT, United Kindom

[b] School of Metallurgy and Materials, The University of Birmingham, Elms Rd, Birmingham B15 2SE, United Kingdom

[c] The Faraday Institution, Quad One, Harwell Campus, Didcot OX11 0RA, United Kingdom

Dr. P. Zhu
Email: p.zhu@bham.ac.uk
orcid.org/0000-0002-0197-7054

Prof. P. R. Slater
Email: P.R.SLATER@bham.ac.uk
orcid.org/0000-0002-6280-7673

*Prof. E. Kendrick
Email: E.Kendrick@bham.ac.uk
orcid.org/0000-0002-4219-964X

**Abstract**

The ever-increasing number of spent lithium-ion batteries (LIBs) has presented a serious waste-management challenge. Aluminium and copper current collectors are important components in LIBs and take up a weight percentage of more than 15%. Direct reuse of current collectors can effectively reduce LIB waste and provide an alternative renewable source of aluminium and copper. Besides, it also prevents long manufacturing processes and associated energy input and material consumption. However, there is a lack of work on the direct reuse of current collectors in the literature. Herein, aluminium and copper current collectors are reclaimed from commercial spent LIBs with different chemical treatments and successfully reused for $LiNi_{0.6}Mn_{0.2}Co_{0.2}O_2$ cathodes and graphite anodes, respectively. The reclaimed current collectors treated with different processes show different surface compositions and morphology to pristine ones, resulting in distinctive wettability, adhesion and electrical conductivity. The reused current collectors show similar electrochemical performance to the pristine one at low C rates, while extra caution should be taken at high C rates for aluminium current collectors due to relatively low contact conductivity. This work provides substantial evidence that the direct reuse of aluminium and copper current collectors is possible and highlights the importance of the surface morphology of current collectors.

**Introduction**

Since the first commercial Lithium-ion battery (LIB) was released by Sony in 1991, the past three decades have witnessed an explosive growth of LIBs.[1] The development of LIBs provides a green technology for energy storage, which can help to address the global climate crisis and build a more sustainable society.[2-3] Nevertheless, LIBs contain many kinds of hazardous and valuable metals,[4] presenting a growing waste-management challenge when reaching the end of their lives,[5-6] which is approximately 10 years for commercial LIBs in electric vehicles (EVs).[7] The volume of spent LIBs was around 260,000 tonnes in 2019 and is predicted to increase to 1.4 million tonnes in 2030.[8-9] It was reported that only 120,000 tonnes of spent LIBs were recycled in 2019,[10] lower than the EU target of 70% recycling rate.[11] Therefore, how to deal with ever-increasing LIB waste has become an urgent issue for our society.

Recycling spent LIBs has been regarded as a feasible and efficient strategy to address the challenge of LIB waste.[12] Most recycling studies have been focusing on electrode active materials,[13-16] while less attention has been given to inactive components, for example, current collectors, separators, electrolytes, and cases. Current collectors are normally aluminium and copper foils and take up more than 15% of the weight of LIBs.[17] Recycling current collectors can effectively reduce LIB waste and provide a large secondary source for Al and Cu. In LIB recycling processes which use hydrometallurgy processes to recycle the active material, current collectors are normally isolated from the black mass during the pre-treatment processes, and are reclaimed by comminution and sieving to produce Al and Cu concentrates which can enter Cu and Al refining processes. Pyrometallurgical recycling processes will consume Al as a reducing agent, and will produce a matte of Cu along with Ni and Co, which will require a further separation step involving hydrometallurgically processes.[13, 18-19] The obtained Al and Cu need further melting, casting, and rolling before reusing, which requires not only high energy input and labour cost but also environmental costs, such as $CO_2$ emissions, freshwater ecotoxicity and marine eutrophication.[20] Furthermore, Al and Cu may contaminate subsequent waste streams and lower the recovery rate of other valuable metals, such as Co and Li. Complete removal of Al and Cu is therefore beneficial for improving subsequent recovery efficiency with fewer purification processes.[21-22] With these issues in mind, direct reuse of current collectors provides an even better option than conventional recycling, which cannot only solve

LIB waste problems but also skip various recycling processes, retaining the embedded environmental and economic benefits. Some publications have proposed the possibility of direct reuse of current collectors in the literature,[23-24] however, there is a lack of experimental work on the direct reuse of current collectors to date.

Current collectors can strongly impact electrode performance although they do not participate in reactions during cell cycling.[17] Al and Cu current collectors transport electrons generated at electrodes to power external circuits and the connection between current collectors and electrodes is crucial for maintaining good electrical contact. The surface composition and morphology of current collectors can be changed by corrosion after long LIB cycling,[25-26] or during the reclamation process, particularly when alkaline or acidic solutions are used.[27-28] The change in the current collector surface can influence the contact between the current collector and electrodes, which in turn affects electrode properties and performance.[17, 29] Therefore, it is necessary to understand the effects of current collector surface conditions when reusing current collectors.

Herein, Al and Cu current collectors are reclaimed from commercial spent LIB cells and directly reused with $LiNi_{0.6}Mn_{0.2}Co_{0.2}O_2$ (NMC622) cathodes and graphite anodes, respectively. Delamination mechanisms of the used Al and Cu current collectors with different treatments and the corresponding surface composition and morphology are investigated. The effects of Al and Cu surface morphology on the wetting of electrode slurries, adhesion strength and electrical conductivity are elucidated. Furthermore, reclaimed Al and Cu current collectors are tested in LIB half cells and compared with pristine Al and Cu current collectors to verify the feasibility of directly reusing the current collectors. We believe that this work can provide substantial evidence for the direct reuse of current collectors and also an insight into the effect of the surface morphology of current collectors on LIB performance.

**Methods**

**Cell disassembly**

End-of-life (EoL) LIB cells were automotive pouch cells from 1st generation Nissan Leaf. The cathode is a mixture of 75% lithium manganese oxide spinel (LMO) and 25% Nickel Cobalt Aluminium oxide (NCA) on an Al current collector, while the anode is graphite on a Cu current collector. The cells were firstly discharged to 2.5 V before dismantling. The discharged EoL cells were then transferred to an Argon-filled glovebox and manually opened using a ceramic scalpel. The components were then separated to obtain cathode coatings on Al foils and anode coatings on Cu foils. The cathode and anode coatings were washed with anhydrous Dimethyl carbonate (DMC) and dried before subsequent treatments. A schematic illustration of cell disassembly is shown in Fig. 1a. More detailed information was reported in.[30]

**Current collector reclamation**

The delamination of cathodic/anodic coatings from Al/Cu foils follows different procedures, as illustrated in Fig. 1b. Two different routes were investigated. Cathode sheets were immersed in N-methyl-2-pyrrolidone (NMP) solution at 60 °C overnight to dissolve polyvinylidene fluoride (PVDF) binder. The cathode coating was then washed using an eraser sponge until no obvious black coating can be seen on the surface. The obtained Al foil is named 'washed Al'. Alternatively, cathode sheets were immersed in 0.5 M oxalic acid at 50 °C under sonication (40 Hz, 50 W) for 5 mins. The cathode coatings separated from Al foils during the process since the Al surface is etched by oxalic acid. The obtained Al foil is named 'etched Al'. As for anodes, anode sheets were firstly soaked in distilled water. Graphite coatings quickly separated from Cu foils in a few seconds. Then, the obtained Cu foils were further washed by 3 M HCl at 30 °C for 5 mins to remove surface oxides (washed Cu). The HCl-washed Cu foils were then transferred to 2 M $HNO_3$ at 30 °C for 30 mins to etch the Cu to get rough surface morphology (etched Cu). All foils were rinsed with distilled water and dried at 60°C before use.

**Electrode making**

NMC622 cathodes and graphite anodes were coated on the reclaimed Al and Cu current collectors, respectively. The cathode contains 96 wt% NMC622 (Targray), 2 wt% PVDF (Solvay) and 2 wt% $C_{65}$ (Imerys). The anode contains 95.25 wt% artificial graphite (S360 E3 Artificial Graphite, BTR), 1.5 wt% CMC (BVH8, Ashland), 2.25 wt% SBR (BM451-B, Zeon), and 1 wt% C45 (Imerys). The mixing procedures are detailed in the supplementary information. The mixed slurries were then coated on Al and Cu current collectors with an areal capacity of around 2 mAh/cm$^2$. NMC622 and graphite coatings were initially dried at 80 and 50 °C, respectively, to remove most solvents followed by overnight drying in a vacuum oven at 120 °C. The dried electrodes were calendered to porosities of around 40% for electrochemical tests. Pristine Al and Cu current collectors (Xiamen TMAX) were also used for comparison.

**Contact angle measurement**

The contact angle of electrode slurries on Al and Cu current collectors was measured by a contact angle goniometer (Ossila). Current collectors were cut into a rectangular piece of 20 x 60 mm$^2$ and placed on a glass slide. A volume of around 40 μL of the slurry was dropped on the current collectors using a pipette. The contact angle was measured 30 s after the droplet was dropped on the current collector.

**Adhesion testing**

The adhesion force between current collectors and electrode coating was measured by a 180-degree peel method using a modified Netzsch Kinexus Pro+ Rheometer. A 25 mm-wide piece of double-sided tape was attached to a section of the top surface of the electrode coating, with the free ends of the coating and the tape attached to the upper and lower rheometer geometry, respectively. The two rheometer geometries then vertically move up and down, respectively, to enact a 180-degree peel at a speed of 10 mm/s with axial force measurement from the rheometer. The force was recorded in a region where the axial force measurement was stable and peeling was visually consistent. The experimental setup can be seen in the inserts in Fig. 5. The adhesion force was determined by taking the average value of five samples for each current collector.

**Electrical conductivity measurement**

The electrical conductivity of NMC622/graphite electrodes on Al/Cu current collectors was measured by a four-point probe (Ossila). All samples were cut in a cylindrical shape with a constant diameter of 14.8 mm. The samples were placed with the electrode coating upwards on a glass slide for testing. The probe spacing is 1.27 mm. The maximum applied voltage was 5V and the maximum current was set to be 1mA for NMC622 on Al and 100 mA for graphite on Cu. This was repeated on five samples for each current collector to take an average.

**Electrochemical testing**

2032 type coin cells were constructed for electrochemical testes, with electrodes on the reused or pristine current collectors (14.8 mm in diameter), tri-layer 2025 separator (Celgard, 16 mm in diameter), Li-metal disc (15 mm in diameter) and 70 µL of 1 M $LiPF_6$ in EC: DMC 3:7 v/v + VC 1%wt (PuriEL). After the assembly of coin cells, a formation step composed of two charge–discharge cycles at a C-rate of 0.05 C (1C=175 mAh/g for NMC622, 1C=350 mAh/g for graphite) was conducted within a voltage window of 2.75 to 4.3 V vs. $Li/Li^+$ for NMC622 half-cells and 0.005 to 1.5 V vs. $Li/Li^+$ for graphite half-cells, using a BCS-805 Biologic battery cycler (Biologic, France). The NMC622 half cells were then charged and discharged at different C rates of 0.1 to 5 for 5 cycles at each C rate. The graphite half cells were discharged and charged at different C rates from 0.1 to 10 for 5 cycles at each C rate. The capacity fade was tested for 50 cycles.

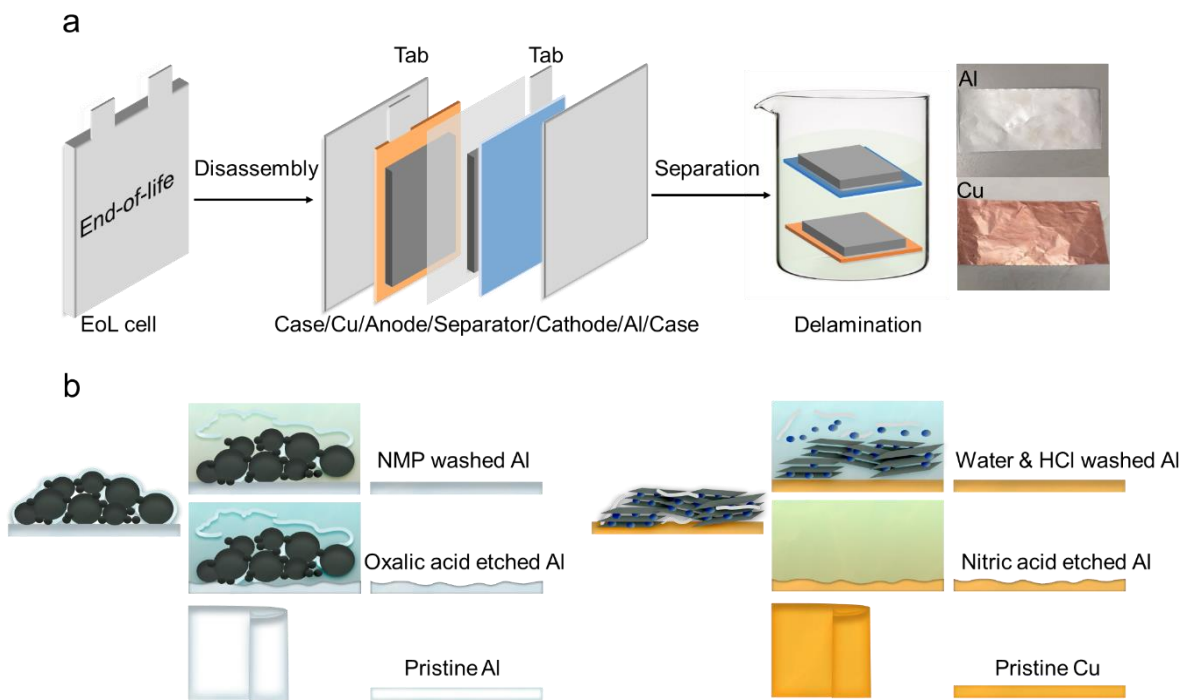

Fig. 1 a) reclamation of Al and Cu current collectors from end-of-life lithium-ion battery cells; b) different treatment conditions for current collectors

## 3. Result and discussion

### 3.1. Surface composition and morphology

The mechanism of the delamination of the electrodes from current collectors differs under different treatments, as illustrated in Fig. 1b. For Al, NMP can dissolve PVDF binder but does not react with Al current collectors.[23, 31] NMP soaking can therefore effectively remove the PVDF binder, and subsequent scrubbing can therefore easily remove the cathode from the surface of Al current collectors. Oxalic acid can react with Al foil and generates hydrogen.[21] The hydrogen bubbles generated at the Al/cathode interface can exert a force to separate the Al current collector and the cathode.[32] The dissolution of Al in oxalic acid is evidenced by the ICP-OES test in Table S1. The delamination of the anode from Cu current collectors mainly relies on lithium leaching and hydrogen bubbles generated at the Cu/graphite interface, exerting a force to separate the Cu current collector and the anode. The water solution becomes very basic after lithium leaching and can oxidise the Cu current collector, as evidenced by a trace amount of Cu in the solution in Table S2. Dilute HCl then removes the surface oxides but does not dissolve the copper foil.[33] Further etching with 2M $HNO_3$ roughens the surface of the Cu current collector.[34]

Fig. 2 a and c show survey-scan XPS spectra of Al and Cu current collectors. All of the three Al current collectors have very similar compositions on the surface except the washed and etched Al have more fluorine (~686 eV) on their surfaces. A zoom-in XPS spectrum for F 1s is shown in Fig. S1. The appearance of fluorine can be attributed to incomplete removal of binder or a thin film of $AlF_3$ formed on the Al surface during LIB cycling.[35] After oxalic acid etching, the peak intensity of fluorine was reduced, indicating a reduced quantity. Fig. 2b further displays a high-resolution XPS spectrum for Al 2p. Two Al 2p peaks at 72.9 and 75.2 eV are assigned to Al and Al oxide, respectively.[36] The pristine Al has a thick surface oxide layer which is formed during production processes, while the etched Al has a thinner Al oxide layer due to the action of the oxalic acid and a thin layer of Al oxide is formed. The washed Al has a much weaker signal of Al oxide and Al than the pristine and etched Al, which is indirect evidence that the surface is covered by some residuals. As for Cu, all three Cu current collectors contain very similar elements, mainly including Cu, O and C, as illustrated in Fig. 2c. Fig. 2d further demonstrates that the washed and etched Cu has a much thinner oxide layer than the pristine Cu because HCl washing can easily remove the oxide layer and a thin oxide layer is subsequently formed when exposed in air.

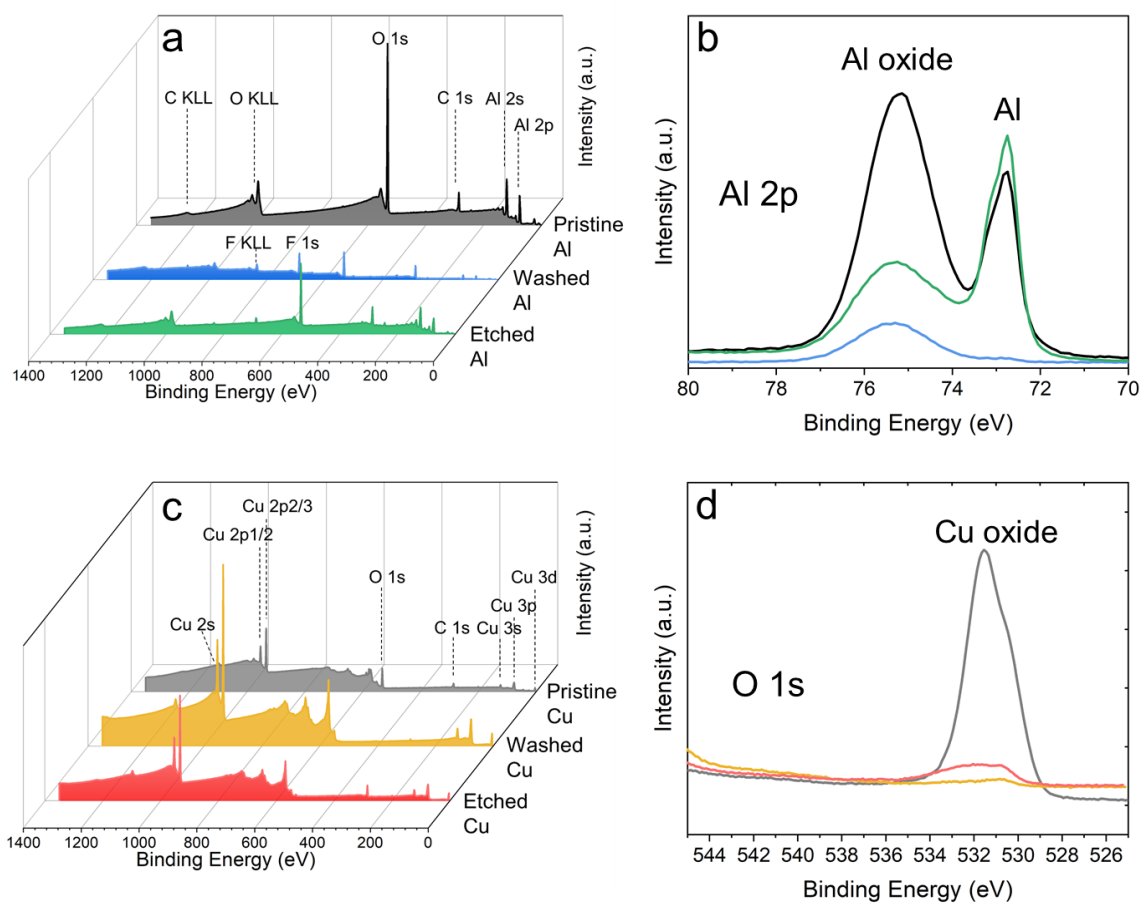

Fig. 2 Survey-scan XPS spectra for a) Al current collectors and c) Cu current collectors, high-resolution XPS spectra for b) Al 2p and d) O 1s.

Fig. 3 shows the surface morphology of the Al and Cu current collectors under different treatments. To facilitate comparison, pristine Al and Cu current collectors are also shown in Fig. 3a and d, respectively. For Al current collectors, the washed Al shows some features including craters, and rolling or calendering traces on the surface, giving a slightly rougher surface than the pristine Al.[37] Given that NMP can only dissolve PVDF and does not react with Al, the washed Al is expected to maintain its original surface morphology after treatment. Etched Al shows a much rougher surface than the washed Al and pristine Al. The surface of the etched Al is full of etched pits and the rolling trace is almost invisible in Fig. 3c due to the reaction between oxalic acid and Al. The size of the etched pits is up to 7 microns in diameter (Fig. S2a). As for Cu current collectors, the washed Cu shows a relatively flat and feature-free surface except for some corrosion pits and cracks (Fig. 3e) as a result of oxidative dissolution during cycling and/or over-discharging prior to cell disassembly.[38] The size of the corrosion pits on the washed Cu is at a submicron scale (Fig. S2b). Etched Cu shows the roughest surface among all Cu current collectors, with numerous etched grooves formed during HNO$_3$ soaking, as shown in Fig. 3f. The width of the etched grooves is below 1 μm (Fig. S2c), much smaller than the etched features on the Al current collector. It is worth mentioning that the corrosion pits and cracks on the washed Cu (Fig. 2e) facilitate further HNO$_3$ etching. A reference group that a pristine Cu with a flat surface cannot achieve the same surface roughness even with the same HNO$_3$ treatment (Fig. S2d).

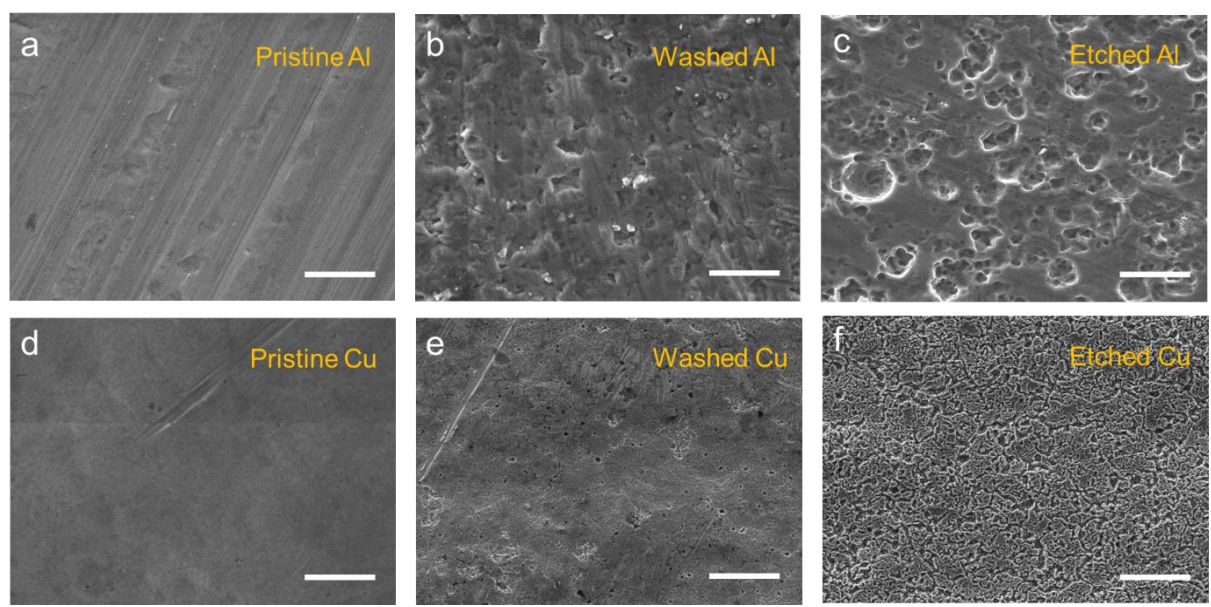

Fig. 3 SEM micrographs of Al and Cu current collectors. a) pristine Al, b) washed Al, c) etched Al, d) pristine Cu, e) washed Cu and f) etched Cu (scale bar: 10 μm).

### 3.2. Wettability

The wettability can be significantly affected by surface morphology.[39-40] Fig. 4 plots the contact angle of NMC622 and graphite electrode slurries on Al and Cu current collectors. The contact angle is an indicator of wettability, and a lower contact angle indicates better wettability. The contact angles of NMC622 slurry on the pristine, washed, etched Al current collectors are 47.38, 44.02, 39.66 degrees, respectively, suggesting that the etched Al exhibits the best wettability, and the pristine Al exhibits the worst wettability. The wettability increases with increasing surface roughness. This is because the surface features on the washed and etched Al current collectors are several microns-size, which are big enough to allow NMC622 slurry to enter, giving rise to an enhanced contact area and therefore better wettability (as evidenced by the cross-section view in Fig. S3i).[40] On the contrary, the reverse is true for the Cu current collectors, where the etched Cu current collector shows the largest contact angle of 102.4 degrees, followed by the washed Cu (87.04 degrees) and pristine Cu (73.94 degrees), revealing that the pristine Cu is the best and the etched Cu is the worst in terms of wettability. In this case, the holes and grooves on the washed and etched Cu are at a submicron scale, which can easily trap air and prevent the graphite slurry from entering the surface features. The surface of the washed and etched Cu actually becomes a Cu-air hydrophobic composite surface and therefore leads to worse wettability.[39] It should be noted that different electrode ink solution systems for making NMC622 and graphite slurries are also a reason for the different wettability performance of Al and Cu current collectors.

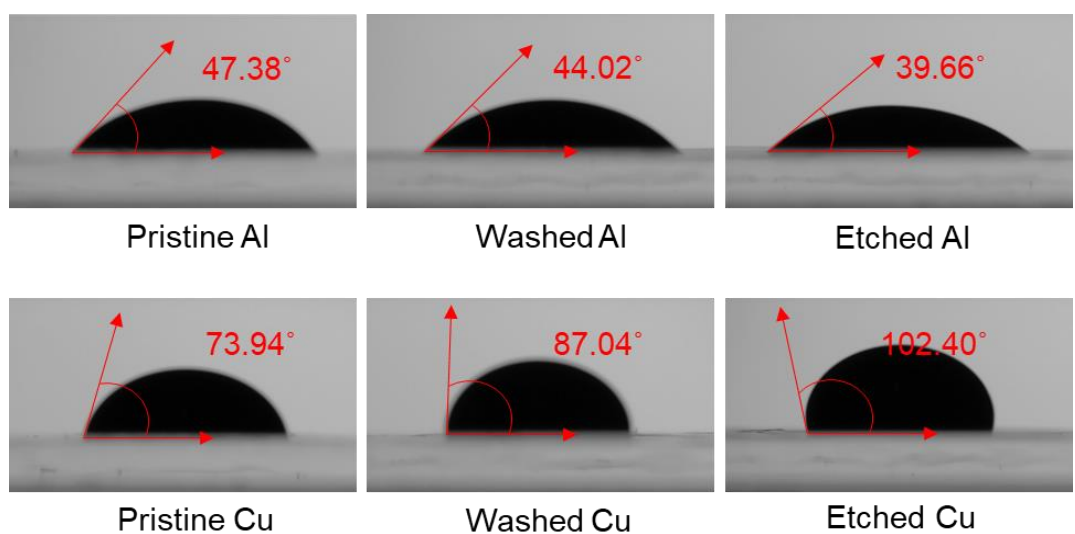

Fig. 4 Wettability of Al and Cu current collectors under different treatments. Contact angle of NMC622 slurry on pristine, washed and etched Al (top), graphite slurry on pristine, washed and etched Cu (bottom).

### 3.3. Adhesion strength

Fig. 5 displays the adhesion force at the interfaces of Al/NMC622 and Cu/graphite measured by the 180-degree peel off test. The electrodes are completely peeled off from the current collectors after the test, as demonstrated in the inserts in Fig. 5. For Al current collectors, the etched Al shows the highest adhesion force of 79.57 N/m, about twice as much as the washed Al (41.71 N/m) and more than five times higher than the pristine Al (15.26 N/m). It is interesting to see that the adhesion force decreases linearly with increasing contact angle for Al (also shown in Table S3), suggesting a strong correlation between wettability and adhesion. The adhesion force between Al current collectors and NMC622 electrodes is directly related to the binder distribution.[41] Fig. S3 shows the image and elemental map of the cross-section of NMC622 on Al current collectors. The distribution of fluorine represents the distribution of PVDF binder. The pristine Al shows an obvious binder distribution gradient, with more PVDF located at the top of the NMC622 electrode, which decreases on moving closer to Al. At the Al/NMC622 interface, the quantity of fluorine is very low, indicating little binder exists at the interface and therefore leads to low adhesion. The low content of PVDF binder at the interfaces between the pristine Al and NMC622 electrode can be directly observed from the cross-section image in Fig. S3a. Though the washed Al shows a slightly more uniform PVDF distribution than the pristine Al, a distribution gradient is still visible, with less PVDF at the washed Al/NMC622 interface. The nonuniform binder distribution is a common issue with electrode manufacturing that results from binder migration during electrode drying and has been systematically reported in a recent review.[42] By contrast, the etched Al shows a more uniform distribution of PVDF throughout the cross-section, with a high PVDF content even at the etched Al/NMC622 interface. The uniform PVDF distribution on the etched Al could be ascribed to the rough Al surface which is easily wetted by NMC622 slurry and traps more PVDF binder at the interface, significantly alleviating binder migration and resulting in higher adhesion strength. It can be directly observed in Fig. S3i that substantial amount of PVDF binder is located at the etched Al/NMC622 interface. It is also worth mentioning that carbon and fluorine show almost the same distribution in Fig. S3, suggesting a high affinity of the carbon black to the PVDF to easily clump together to form a well-known carbon binder domain (CBD) [7].

Unlike the Al current collectors, the three Cu current collectors show very similar adhesion forces, with the etched Cu contributing to slightly high adhesion forces of 2.67 N/m, followed by the washed Cu (2.47 N/m) and the pristine Cu (2.26 N/m). The adhesion force between Cu current collectors and graphite electrodes can also be understood by the binder distribution. Fig. S4 show the image and elemental map of the cross-section of graphite electrodes on Cu current collectors. The distribution of sodium is an indicator of the distribution of CMC. For all Cu current collectors, the CMC distributes evenly within the graphite electrode but with little at the Cu/graphite interfaces, as demonstrated in Fig. S4 d, h, l. The more uniform distribution of the CMC than PVDF might be ascribed to the lower initial drying temperature (50 °C for the CMC and 80 °C for the PVDF). The low binder content at the interfaces mainly results from the poor wettability of the aqueous slurry on the Cu surface. In this respect, it may be of future interest to introduce surfactants, which can be used to decrease the surface tension of electrode slurries and enhance wettability.[43] Apart from the poor wettability, the shape of graphite flakes also influences the stacking and may affect the binder distribution and adhesion.

The magnitude of the adhesion forces between Al/NMC622 and Cu/graphite is significantly different. Comparing Fig. 5a and b, the adhesion at the Al/NMC622 interface is one order of magnitude higher than that at the Cu/graphite interface. It is suspected the difference in adhesion force comes from the different binders being used. To facilitate a direct comparison of the adhesion strength of PVDF and CMC/SBR, the CMC/SBR was replaced with the same weight percentage of PVDF (3.75%) to make the graphite slurry and was coated onto the same Cu current collectors. By using PVDF binder, the adhesion forces increase to 20.34, 69.82 and 101.06 N/m for the pristine, washed and etched Cu (Fig. S5), respectively, about the same level as the Al current collectors, thus indicating that PVDF is a much stronger binder than CMC/SBR in terms of adhesion.

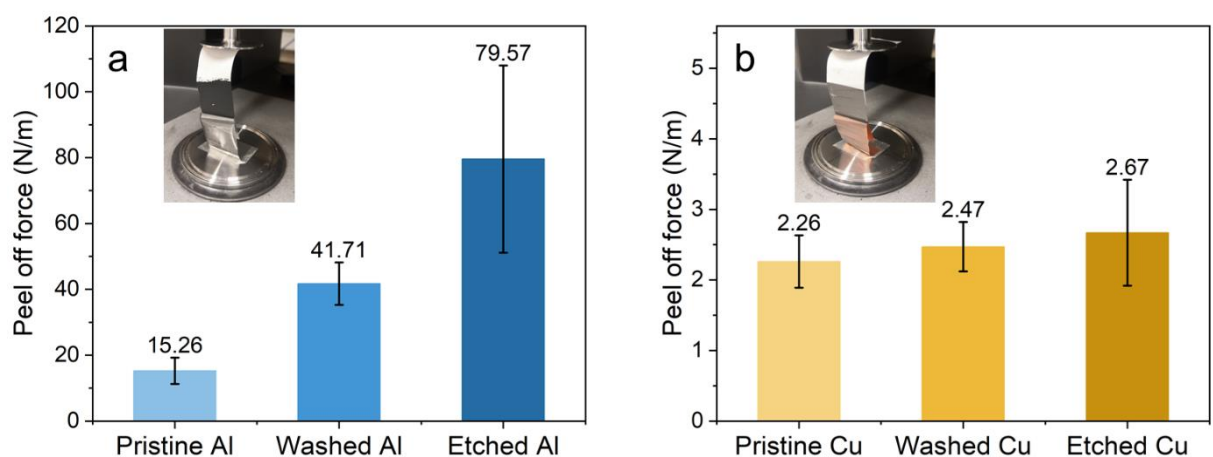

Fig. 5 Adhesion strength between electrodes and current collectors. Peel off force between a) Al current collectors and NMC622 electrode and b) Cu current collectors and graphite electrode.

### 3.4. Electrical conductivity

Fig. 6 plots the electrical conductivity of NMC622 electrodes on Al current collectors and graphite electrodes on Cu current collectors measured by a four-point probe test. The electrical conductivities of NMC622 electrodes on the pristine, washed, etched Al current collectors are 49.07, 20.98, 18.37 S/m, respectively. The electrical conductivity decreases with increasing Al surface roughness, which is opposite to the adhesion strength. This is likely due to Al current collectors with a rougher surface trapping more PVDF binder at the Al/NMC622 interface and blocking the electrical conduction pathway,[44] reducing the contact conductivity. Thus, the rougher the Al surface, the lower the electrical conductivity. The electrical conductivities of graphite electrodes on the pristine, washed, etched Cu current collectors are 1.18, 1.12, 1.20 x $10^6$ S/m, respectively. The effect of the different treatments of the Cu current collectors on electrical conductivity is therefore negligible. As discussed in section 3.3, Cu surface morphology does not affect the distributions of CMC binder and carbon very much. Thus, the electrical conductivities of graphite anodes on three different Cu current collectors are very similar.

The electrical conductivity of NMC622 electrodes on Al current collectors is five orders of magnitude lower than that of graphite electrodes on Cu current collectors. The main reason is that graphite is an electrically conductive material but NMC622 has very poor electrical conductivity.[45-46] In addition, as noted above, the binder can also affect electrical conductivity. Fig. S4 shows that the electrical conductivities of graphite electrodes made with 3.75% PVDF and coated on the pristine, washed, etched Cu current collectors are 2.70, 1.89, 2.31 x $10^5$ S/m, respectively. By replacing CMC/SBR with the same amount of PVDF, the electrical conductivity of graphite electrodes on Cu current collectors can be reduced by about five times, indicating that CMC/SBR is better than PVDF in terms of maximising electrode electrical conductivity.

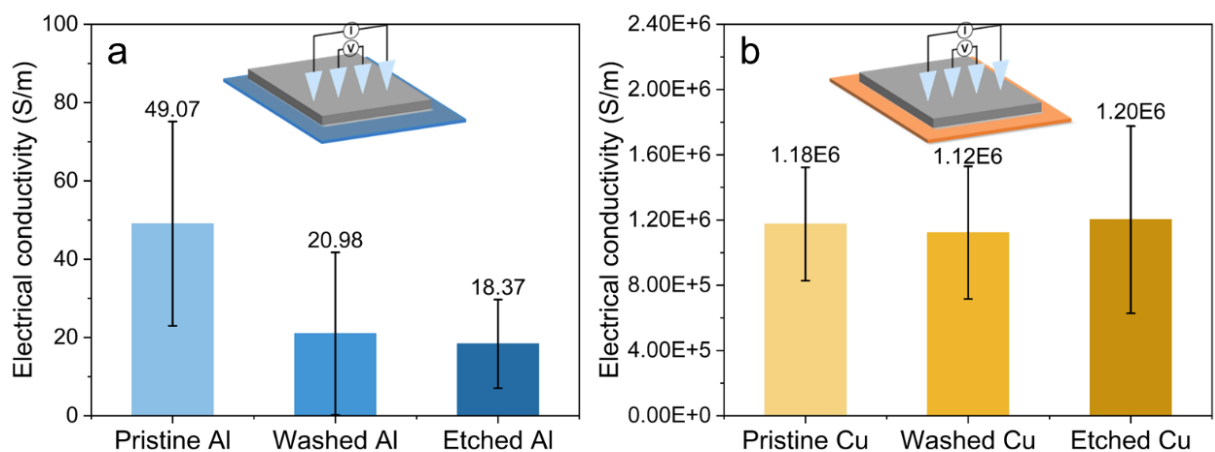

Fig. 6 Four-point probe test for electrodes on current collectors. Electrical conductivity of a) NMC622 electrodes on Al current collectors, b) graphite electrodes on Cu current collectors.

**3.5. Electrochemical testing**

Reclaimed Al and Cu current collectors were evaluated with NMC622 and graphite electrodes in LIB half cells, respectively, and compared with the pristine current collectors to verify the feasibility of direct reusing current collectors. Fig. 7a shows the rate capability of NMC622 electrodes on Al current collectors. At 0.1C, NMC622 electrodes on the pristine, washed and etched Al current collectors deliver similar capacities of around 170 mAh/g which is very close to the theoretical value of 175 mAh/g, indicating nearly full utilisation of the active material.[47-48] In the C rate range from 0.1 to 0.5C, the capacities of NMC622 electrodes on all Al current collectors are similar. As the C rate further increases to 1, 2 and 5C, NMC622 on the pristine Al delivers higher capacities than the counterpart on the washed and etched Al. At 5C, NMC622 on the pristine Al delivers a capacity in the range of 80-100 mAh/g, while NMC622 on the washed and etched Al delivers low specific capacities, less than 15 mAh/g. This is attributed to the NMC622 electrode on the pristine Al having the highest electrical conductivity. As the C rate increases from 0.1 to 1C and even higher, the current density for charging and discharging accordingly increases by ten times and even higher, thus the electrical conductivity becomes a limiting factor. A higher electrical conductivity contributes to a lower voltage drop and therefore a higher capacity. Fig. 7b shows that the NMC622 electrodes on all the pristine, washed and etched Al current collectors have stable cycling over 100 cycles at 0.2C, indicating that the reused Al current collectors have good stability.

For the Cu current collectors, the graphite electrodes on the pristine, washed, etched Cu current collectors deliver similar capacities at a wide range of C rates from 0.1 to 10C, as shown in Fig. 7c. The highest capacity achieved is about 350 mAh/g at 0.1C. Considering that the electrical conductivity of graphite electrodes on Cu current collectors is five orders of magnitude higher than that of NMC622 electrodes on Al current collectors, the electrical conductivity should not be a limiting factor for graphite electrodes even at high C rates. Fig. 7d further displays that the capacity of the graphite electrodes on different Cu current collectors is stable after 100 cycles at 0.2C. Therefore, both the washed and etched Cu can be directly reused and possibly replace pristine Cu current collectors.

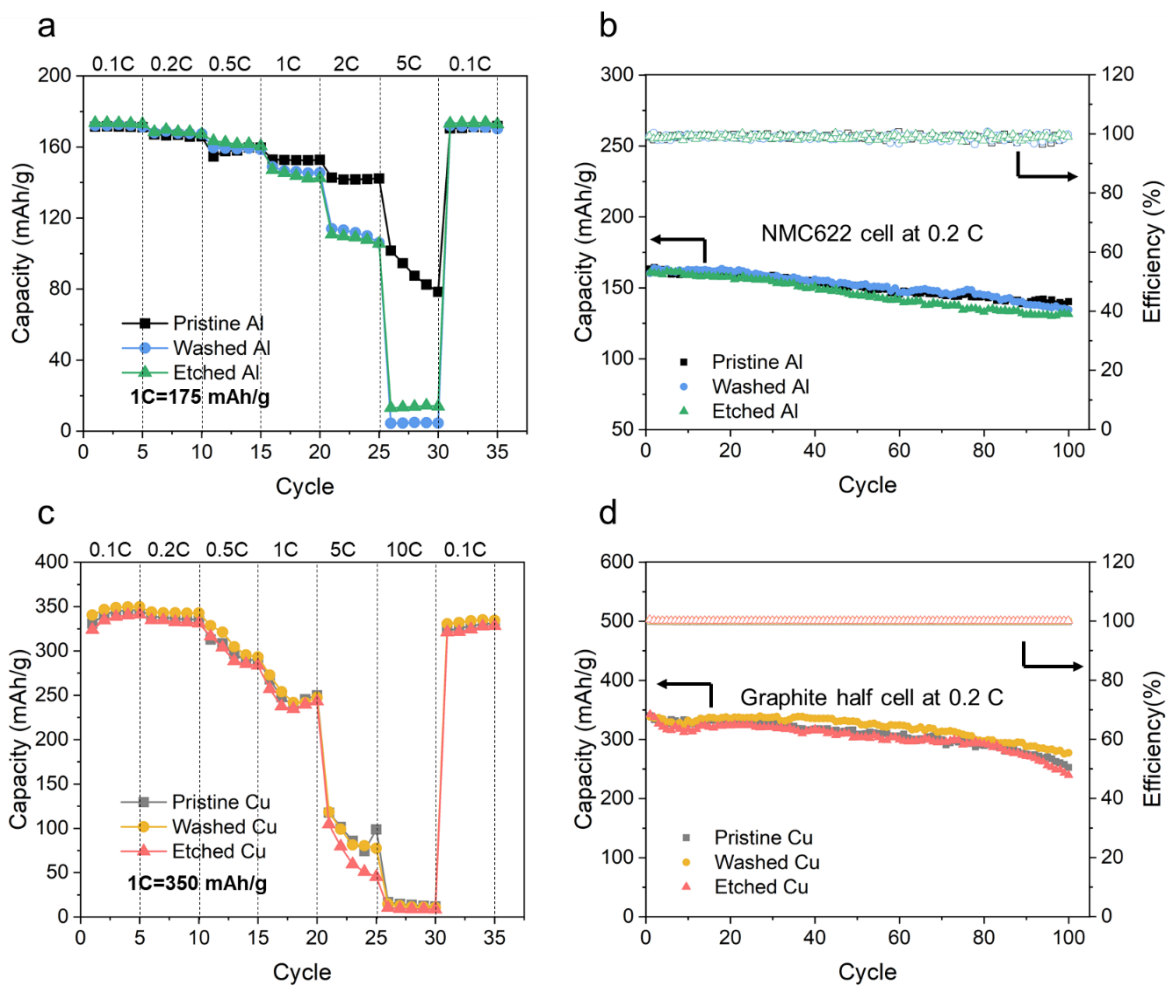

Fig. 7 Electrochemical performance Al and Cu current collectors in half cells. Rate capability of a) NMC622 on Al current collectors and c) graphite on Cu current

collectors; cyclability of b) NMC622 on Al current collectors at 0.2C and d) graphite on Cu current collectors at 1

## 4. Conclusions

Al and Cu current collectors reclaimed from spent commercial LIBs have been successfully directly reused with some simple treatments, including NMP washing and oxalic acid etching for Al, water soaking with subsequent HCl washing and optional $HNO_3$ etching for Cu. As summarised in Table S3, the surface composition of current collectors changes slightly but the surface morphology varies distinctively with different treatments. The roughness of the surface morphology of both Al and Cu follows the same order, etched > washed > pristine. The reused Al with higher surface roughness results in better wetting of NMC622 slurry, enhanced adhesion at the Al/NMC622 interface, more uniform PVDF binder distribution, but reduced electrical conductivity. In contrast to Al, the effects of Cu surface morphology on the adhesion at Cu/graphite interface and electrical conductivity, CMC binder distribution are not pronounced. The reused Al current collectors deliver very similar capacities at 0.1 – 1C but lower capacities at higher C rates when compared with the pristine Al, while the reused Cu current collectors exhibit almost the same capacities as the pristine Cu at a wide C rate range from 0.1-10C. This work details substantial evidence of direct reusing of current collectors, providing an alternative renewable source of current collectors and preventing traditional long manufacturing processes and associated energy input and material consumption for new current collectors.

Although the feasibility of reusing Al and Cu current collectors has been verified and the effect of surface morphology has been investigated in this work, future effort is still necessary, particularly for two areas below: 1) Conflict between interfacial adhesion and contact conductivity. Considering that polymeric binders are normally non-conductive, higher adhesion strength requires more binders at the current collector/electrode interface, which inevitably reduces the contact conductivity. Developing new binders with improved adhesive performance or with good electrical conductivity is therefore a key target. 2) Scaling up for application in industry. Current work was conducted at a lab scale, the scalability needs to be further demonstrated. Furthermore, it is easy to cause wrinkles on current collectors during the reclamation processes, and so strategies to maintain flatness while treating a large quantity of current collectors will be a challenge.

**Conflict of Interest**

The authors declare no conflict of interest.

**Acknowledgements**

This work is supported by the Faraday Institution-funded Nextrode (FIRG015). CATMAT (FIRG016) and ReLIB (FIRG27) projects. The x-ray photoelectron (XPS) data collection was performed at the EPSRC National Facility for XPS ("HarwellXPS"), operated by Cardiff University and UCL, under Contract No. PR16195.

**Supplymentary Information**

**Direct Reuse of Aluminium and Copper Current Collectors from Spent Lithium-ion Batteries**

Pengcheng Zhu [a, c], Elizabeth H. Driscoll [b, c], Bo Dong [a, c], Roberto Sommerville [b, c], Anton Zorin [b, c], Peter R. Slater [a, c], Emma Kendrick [b, c] *

[a] School of Chemistry, The University of Birmingham, Birmingham, B15 2TT, United Kindom

[b] School of Metallurgy and Materials, The University of Birmingham, Elms Rd, Birmingham B15 2SE, United Kingdom

[c] The Faraday Institution, Quad One, Harwell Campus, Didcot OX11 0RA, United Kingdom

**Electrode slurry mixing**

For cathode mixing, PVDF was pre-dissolved in NMP to make a PVDF solution with a concentration of 8 wt%. Half the 8 wt% PVDF solution was mixed with $C_{65}$ using a THINKY mixer (ARE-20, Intertronics) at 500 rpm for 1min and 2000 for 5 mins. NMC622 and the other half of the PVDF solution were added to the mixture and mixed again at 500 rpm for 1 min and 2000 rpm for 10 mins. The mixture was subsequently degassed in the THINKY mixer at 2200 rpm for 3 mins. The obtained slurry was homogenous and had a solid content of around 60%. For anode mixing, CMC was pre-dissolved in distilled water to make CMC solution with a concentration of 1.5 wt%. Half of the 1.5 wt% CMC solution was firstly mixed with $C_{45}$ in THINKY mixer at 500 rpm for 1 min and 2000 rpm for 5 mins. Graphite and the remaining CMC solution were added and mixed again at 500 rpm for 1 min and 2000 rpm for 10 mins. The mixture was then degassed in the THINKY mixer at 2200 rpm for 3 mins. 2.25 wt% SBR solution with a concentration of 40% was added at the end and mixed at 500 rpm for 5 mins. The prepared graphite slurry was homogenous and had a solid content of around 50%.

**Inductively coupled plasma - optical emission spectrometry (ICP-OES) analysis**

10 $cm^2$ discs were cut from each current collector using a calibrated James-Heal sample cutter (James-Heal, UK). These samples were digested for 45 minutes with an average temperature of 170 ℃ in 10 mL of Aqua Regia (4:1 ratio of HCl : $HNO_3$ (70% aqueous)) using a microwave digester (Anton-Paar, Austria). The resulting mixtures were filtered using quantitative filter paper (Fisher Scientific, UK) and made up to 250 mL in volumetric flasks, using distilled water (all glassware was washed thrice with distilled water). A sample was taken from this dilution for analysis (100% baseline samples). A 5 mL aliquot was taken from the 250 mL volumetric flask and diluted up to 50 mL in volumetric flasks - again a sample was taken from this dilution (10% samples).

The samples were analysed using an Agilent 5110 ICP-OES with an argon plasma torch (Agilent Technologies, USA). Three repeat measurements were taken for each individual sample by the instrument with a rinse step between each sample. A re-slope of the calibration line was performed every 20 samples using the 0 ppm, 5, 20 and 70 ppm standards. An error of 20% and an $r^2$ value of 0.995 were chosen for the calibration lines.

For the axial standards (0 - 15 ppm), a 50 ppm working solution consisting of 9 elements was prepared from 1000 ppm single element standards (Al, Co, Cu, Fe, Li, Mn, Na, Ni, and P). This working solution was appropriately diluted down to produce the desired standards using volumetric flasks. These standards were acidified using aqua regia.

For the radial standards (20 - 100 ppm), these standards were prepared individually (rather than previous, where a working solution of 50 ppm was used) using 1000 ppm element standards (Al, Co, Cu, Fe, Li, Mn, Na, Ni, and P) to create the desired concentrations. These were also acidified using aqua regia.

The 20 ppm solution was used in the calibration of both the axial and radial measurements.

**X-ray photoelectron spectroscopy analysis**

XPS Analysis was performed using a Thermo NEXSA XPS fitted with a monochromated Al kα X-ray source (1486.7 eV), a spherical sector analyser and 3 multichannel resistive plates, 128 channel delay line detectors. All data was recorded at 19.2W and an X-ray beam size of 400 x 200 μm. Survey scans were recorded at a pass energy of 200 eV, and high-resolution scans were recorded at a pass energy of 40 eV. Electronic charge neutralization was achieved using a Dual-beam low-energy electron/ion source (Thermo Scientific FG-03). Ion gun current = 150 μA. Ion gun voltage = 45 V. All sample data were recorded at a pressure below $10^{-8}$ Torr and a room temperature of 294 K. All Al and Cu current collectors were sputtered for 20 and 30s using Ar 4000+ eV monatomic mode with a raster size of 2x2 mm^2 (etching rate 0.57 nm/s ref. $Ta_2O_5$), respectively, to remove surface contamination. Data were analysed using CasaXPS v2.3.20PR1.0. Peaks were fitted with a Shirley background prior to component analysis. Line shapes of LA(1.53,243) were used to fit components.

**Scanning electrode microscopy with energy-dispersive X-ray (SEM-EDX) spectroscopy analysis**

The surface morphology of the reclaimed and pristine current collectors was investigated by scanning electron microscopy (SEM, Philips XL30 FEG) under an acceleration voltage of 10 kV. Magnifications of 2000x and over were utilised for the observation of surface features on all current collectors. The elemental distributions were measured by energy-dispersive X-ray spectroscopy (EDX, Oxford Inca 300).

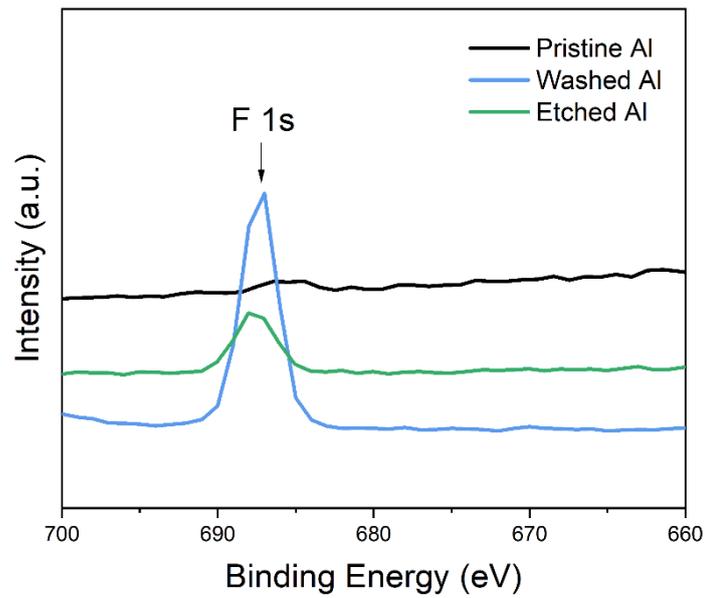

Figure S1. Zoom in XPS spectra for F 1s.

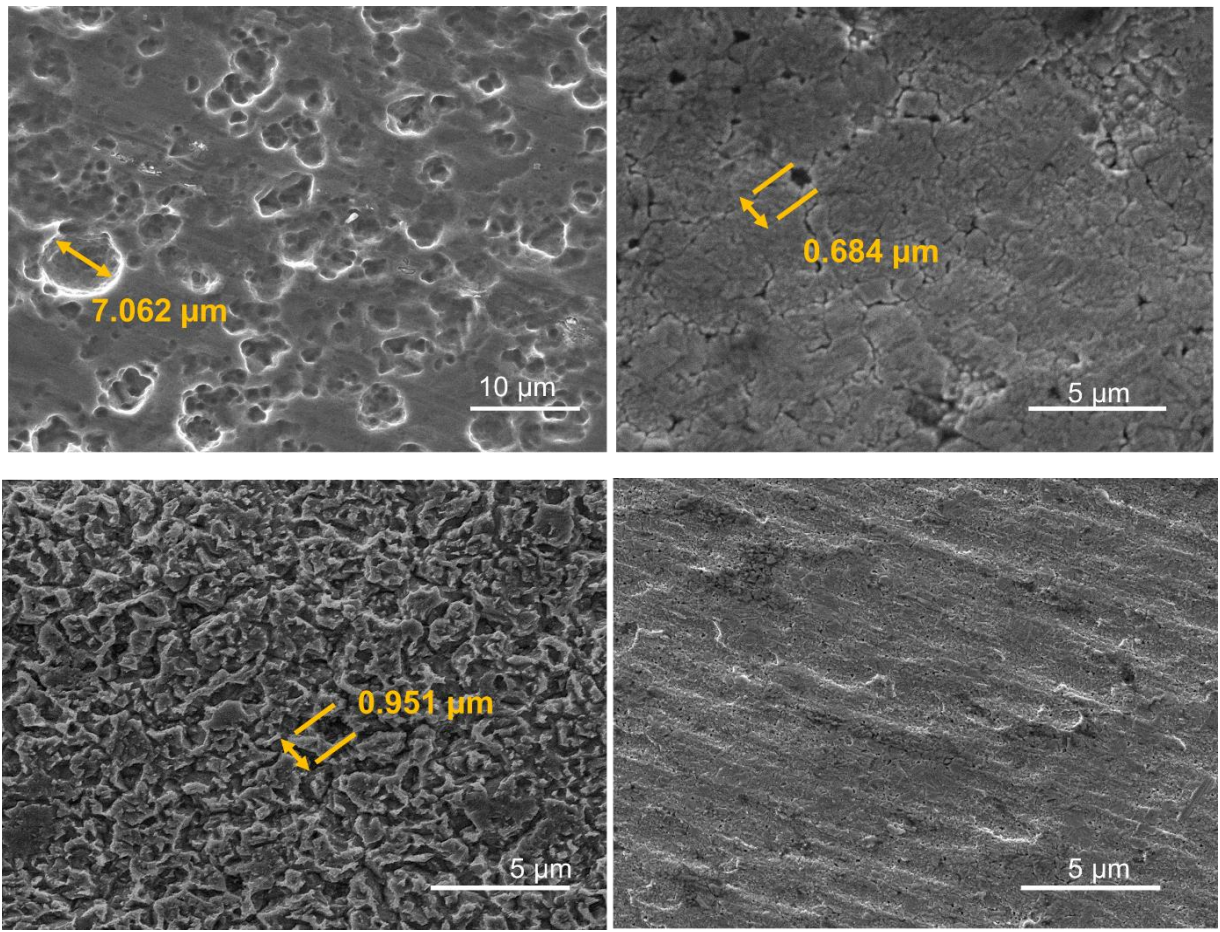

Figure S2. Size of etched pits on etched Al surface measured by Image J.

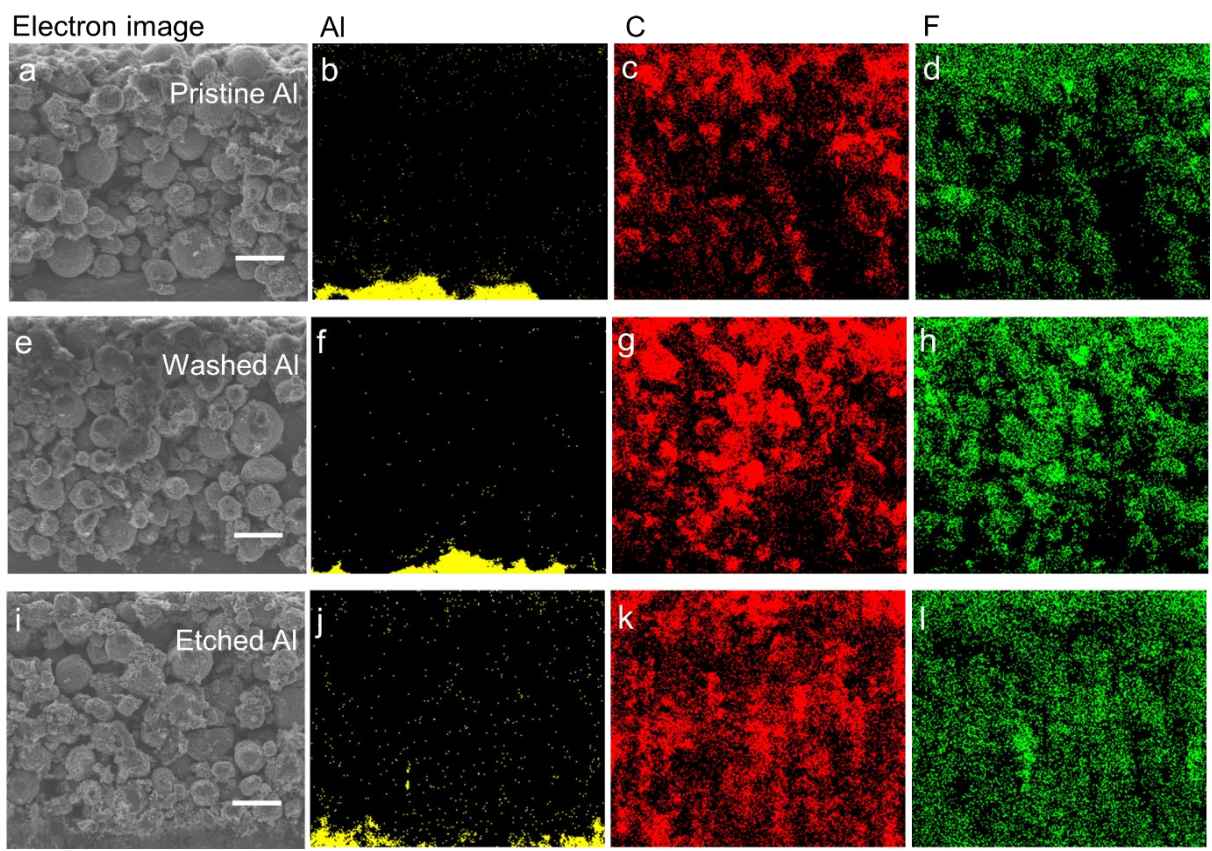

Figure S3 SEM micrographs and EDS mapping of the cross-section of NMC622 electrodes on pristine (a-d), washed (e-h) and etched (i-l) Al current collectors. Micrographs of NMC622 on pristine (a), washed (e) and etched (i) Al current collectors; EDS mapping of Al (b, f, j), carbon (c, g, k) and fluorine distribution (d, h, l) (scale bar: 20 μm).

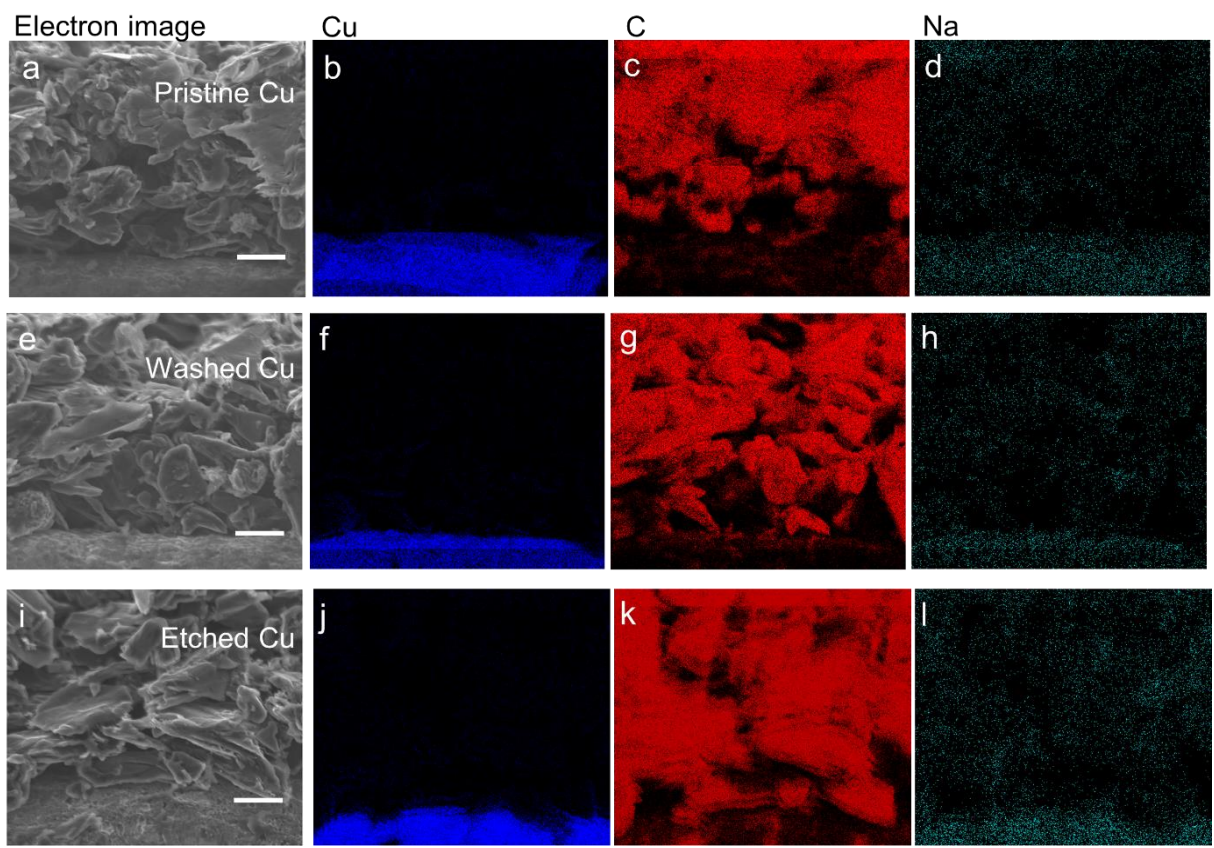

Figure S4 SEM micrographs and EDS mapping of the cross-section of graphite electrodes on pristine (a-d), washed (e-h) and etched (i-l) Cu current collectors. Micrographs of graphite on pristine (a), washed (e) and etched (i) Cu current collectors; EDS mapping of Cu (b, f, j), carbon (c, g, k) and sodium distribution (d, h, l) (scale bar: 20 μm).

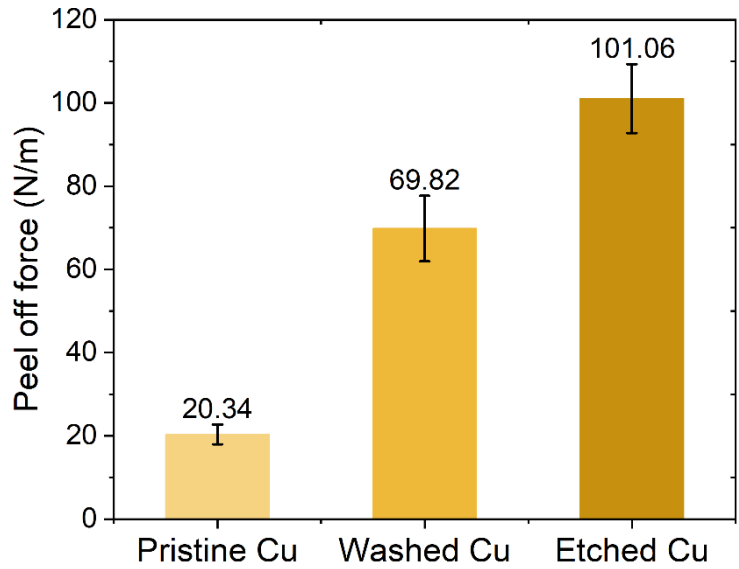

Figure S5. Adhesion force of graphite electrodes made with PVDF binder on Cu current collectors

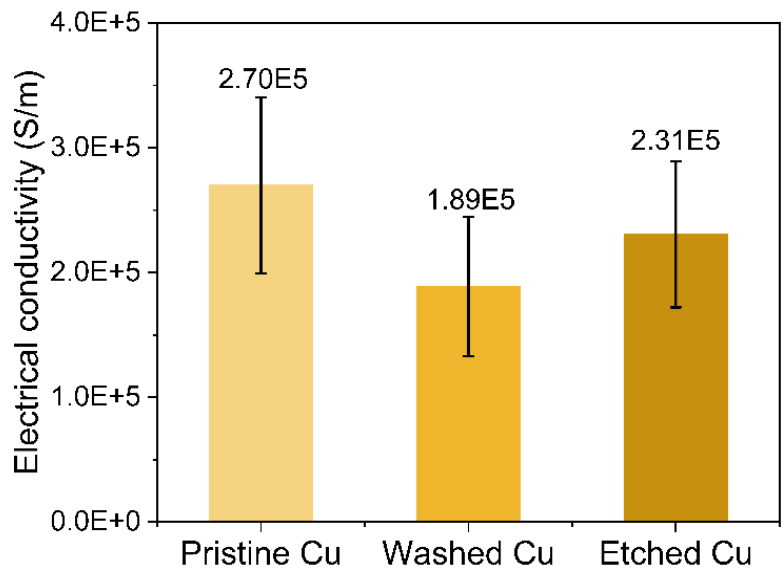

Figure S6. Electrical conductivity of graphite electrodes made with PVDF binder on Cu current collectors

**Table S1: ICP-OES test for Al current collector in oxalic acid**

| Element | Al | Co | Cu | Li | Mn | Ni |
|---|---|---|---|---|---|---|
| Concentration (ppm) | 109.27 | 27.01 | 0 | 26.23 | 347.93 | 109.44 |

**Table S2: ICP-OES test for Cu in water**

| Element | Al | Co | Cu | Li | Mn | Ni | P |
|---|---|---|---|---|---|---|---|
| Concentration (ppm) | 0.04 | 0.03 | 0.04 | 6.23 | 0.07 | 0.04 | 3.28 |

**Table S3: Reused and pristine Al and Cu current collectors**

| Material | Treatment | Surface composition | Surface roughness | Wettability (degree) | Adhesion (N/m) | Electrical conductivity (S/m) | Binder distribution | Capacity @ 0.1 C (mAh/g) | Capacity @ 5 C (mAh/g) |
|---|---|---|---|---|---|---|---|---|---|
| **P_Al** | None | Al oxide | Low | 47.38 | 15.26 | 49.07 | Gradient | 171.48 | 101.64 |
| **W_Al** | NMP | Al oxide & residual PVDF | Medium | 44.02 | 41.71 | 20.98 | Gradient | 172.07 | 4.39 |
| **E_Al** | Oxalic acid | Thin Al oxide | High | 39.66 | 79.57 | 18.37 | Uniform | 173.52 | 13.10 |
| **P_Cu** | None | Cu oxide | Low | 73.94 | 2.26 | 1.18E6 | Uniform | 330.75 | 117.72 |
| **W_Cu** | HCl | Very thin Cu oxide | Medium | 87.04 | 2.47 | 1.12E6 | Uniform | 340.54 | 118.39 |
| **E_Cu** | HCl&HNO$_3$ | Very thin Cu oxide | High | 102.4 | 2.67 | 1.20E6 | Uniform | 323.89 | 104.43 |

Note: P stands for pristine, W stands for washed, E stands for etched